\documentclass[aps,prl,preprint,superscriptaddress,showpacs,floatfix]{revtex4}
\usepackage{graphicx}
\usepackage{dcolumn}
\usepackage{bm}
\usepackage{amsfonts}
\usepackage{amsmath}
\usepackage{amssymb}

\begin{document}
\renewcommand{\thefigure}{\arabic{figure}}

\title{Luther-Emery Phase and Atomic-Density Waves in a Trapped Fermion Gas}

\author{Gao Xianlong}
\affiliation{NEST-CNR-INFM and Scuola Normale Superiore, I-56126 Pisa, Italy}
\author{M. Rizzi}
\affiliation{NEST-CNR-INFM and Scuola Normale Superiore, I-56126 Pisa, Italy}
\author{Marco Polini}
\email{m.polini@sns.it}
\affiliation{NEST-CNR-INFM and Scuola Normale Superiore, I-56126 Pisa, Italy}
\author{Rosario Fazio}
\affiliation{International School for Advanced Studies (SISSA), via Beirut 2-4, I-34014 Trieste, Italy}
\affiliation{NEST-CNR-INFM and Scuola Normale Superiore, I-56126 Pisa, Italy}
\author{M.P. Tosi}
\affiliation{NEST-CNR-INFM and Scuola Normale Superiore, I-56126 Pisa, Italy}
\author{V.L. Campo, Jr.}
\affiliation{Centro Internacional de F\'{\i}sica da Mat\'eria Condensada,
	Universidade de Bras\'{\i}lia, Caixa Postal 04513, 70919-970 Bras\'{\i}lia, Brazil}
\author{K. Capelle}
\affiliation{Departamento de F\'{\i}sica e Inform\'atica,
	Instituto de F\'{\i}sica de S\~ao Carlos,
	Universidade de S\~ao Paulo,
	Caixa Postal 369, 13560-970 S\~ao Carlos, S\~ao Paulo, Brazil}

\begin{abstract}
The Luther-Emery liquid is a state of matter that is predicted to occur in one-dimensional systems of interacting fermions and is characterized by a gapless charge spectrum and a gapped spin spectrum. 
In this Letter we discuss a realization of the Luther-Emery phase in a trapped cold-atom gas.
We study by means of the density-matrix renormalization-group technique 
a two-component atomic Fermi gas with attractive interactions subject to parabolic trapping
inside an optical lattice. We demonstrate how this system exhibits compound phases characterized 
by the coexistence of spin pairing and atomic-density waves. A smooth crossover occurs 
with increasing magnitude of the atom-atom attraction to a state in which tightly bound 
spin-singlet dimers occupy the center of the trap. The existence of atomic-density waves 
could be detected in the elastic contribution to the light-scattering diffraction pattern.
\end{abstract}
\pacs{03.75.Ss,71.10.Fd,71.10.Pm,71.15.Mb}
\maketitle

{\it Introduction} ---  The ability to confine ultracold Bose and Fermi gases 
inside artificial crystals generated by standing-wave laser light fields, {\it i.e.} optical lattices (OLs), 
offers the possibility to create ideally clean and highly tunable strongly interacting quantum many-body systems~\cite{cirac_2003}. The experimental observation by Greiner {\it et al.}~\cite{greiner_2002} of a superfluid-Mott insulator transition in a $3D$ OL loaded with $^{87}{\rm Rb}$ atoms demonstrated the possibility of realizing strongly correlated bosonic systems. A number of experimental results have also been obtained for fermionic systems in OLs~\cite{stoferle_2006}. Cold atoms have been successfully trapped in low-dimensional geometries. A $^{87}{\rm Rb}$ gas has been used to realize experimentally a Tonks-Girardeau gas~\cite{paredes_nature_2004}. The preparation of two-component Fermi gases in a quasi-$1D$ geometry~\cite{moritz_2005} provides a unique possibility to experimentally study phenomena predicted a long time ago for electrons in $1D$. Spin-charge separation in Luttinger liquids is a paradigmatic 
example~\cite{giamarchi}. 

In the presence of attractive interactions fermions are predicted to form a peculiar $1D$ liquid phase 
characterized by a massive spin sector, {\it i.e.} a Luther-Emery liquid~\cite{luther_emery}. 
The gap in the spin sector induces an exponential decay of spin correlations, while singlet superconducting and charge-density wave correlations have a power-law decay~\cite{giamarchi}. 
So far no observation of the Luther-Emery phase has been reported in solid-state electronic systems. In 
Ref.~\cite{LE} it has been shown that an integrable model of two-component interacting 
Fermi gases in a quasi-$1D$ geometry exhibits a smooth crossover between a Luther-Emery 
liquid and a Luttinger liquid of tightly-bound 
bosonic dimers. Although long-range phase coherence is absent in a $1D$ Luther-Emery liquid, antiparallel-spin fermions do pair as they do in a conventional superconductor. Seidel and Lee~\cite{seidel_2004} have shown that the ground-state energy of a system with spin gap and gapless charge degrees of freedom has an exact period of $hc/(2e)$ (corresponding to half a flux quantum) as a function of an applied Aharonov-Bohm flux.

In this Letter we propose that $1D$ Fermi gases in OLs 
be used to study the Luther-Emery phase and the existence of antiparallel-spin pairing in a Luther-Emery liquid. A realistic OL is superposed on a slowly varying harmonic potential that makes the lattice sites inequivalent.
In order to assess the very existence and the nature of the Luther-Emery
phase in realistic cold-atom systems it is thus of fundamental importance to analyze the interplay between attractive interactions and the confining potential. This is not a merely quantitative issue: it is well 
known that in other cases the confinement modifies qualitatively the 
properties of the gas. An example is the coexistence of superfluid (metallic) and Mott-insulating regions 
in bosonic (fermionic) OLs~\cite{jaksch,repulsive,campo_pra_2005,gao_PRB}.

We first show that, in the presence of harmonic confinement, 
a $1D$ Fermi gas with attractive interactions inside an OL 
manifests unambiguous real-space spin pairing, which in turn determines 
the emergence of Atomic-Density Waves (ADWs) 
in the ground-state density profile. We then propose an experiment that can lead to the 
observation of these ADWs.

{\it The inhomogeneous Fermi-Hubbard model} --- 
We consider a two-component Fermi gas with $N$ atoms confined by a 
harmonic potential of strength $V_2$ in a $1D$ lattice with unit lattice constant and $L$ 
lattice sites $i\in[1,L]$. The system is described by an inhomogeneous Fermi-Hubbard Hamiltonian,
\begin{equation}
\label{eq:hubbard}
	{\hat {\cal H}}=-t\sum_{i,\sigma}({\hat c}^{\dagger}_{i\sigma}
	{\hat c}_{i+1\sigma}+{\rm H}.{\rm c}.)
	+U\sum_i\,{\hat n}_{i\uparrow}{\hat n}_{i\downarrow}+V_2\sum_i(i-L/2)^2{\hat n}_i\,.
\end{equation}
Here $t$ is the hopping parameter, $\sigma=\uparrow,\downarrow$ is a pseudospin-$1/2$ degree 
of freedom (hyperfine-state label), $U$ is the on-site Hubbard interaction parameter, 
${\hat n}_i=\sum_\sigma {\hat n}_{i\sigma}=\sum_\sigma {\hat c}^{\dagger}_{i\sigma}{\hat c}_{i\sigma}$ 
and $\langle \sum_i{\hat n}_i\rangle=N$. $L$ is always chosen so that the trap makes the ground-state site occupation $n_i=\langle \Psi_{\rm GS}|{\hat n}_i|\Psi_{\rm GS}\rangle$ go to zero smoothly near the edges of the lattice. The Hamiltonian ${\hat {\cal H}}$ with repulsive interactions ($U>0$) 
has been extensively discussed in the literature~\cite{repulsive,campo_pra_2005,gao_PRB}. 
In this Letter we focus on attractive interactions, {\it i.e.} $U<0$, and 
on the unpolarized case ($N_\uparrow=N_\downarrow$).

In the absence of harmonic confinement ($V_2=0$) ${\hat {\cal H}}$ has been exactly solved by 
Lieb and Wu~\cite{lieb_wu} by means of the Bethe {\it Ansatz} and shown to belong to the Luther-Emery liquid universality class. For $V_2\neq 0$ we calculate 
the ground-state properties of ${\hat {\cal H}}$ in Eq.~(\ref{eq:hubbard}) by resorting 
to the density-matrix renormalization-group 
(DMRG) method~\cite{dmrg}, which provides a practically exact solution for any value of $U/t$. 
Motivated by the recent interest in the development of 
density-functional schemes for strongly correlated $1D$ systems~\cite{campo_pra_2005,gao_PRB,lima_prl_2003,burke_2004,superlattice,disorder} we also use, in parallel to DMRG, a lattice density-functional scheme based on the Lieb-Wu solution for $V_2=0$. These calculations employ the Bethe-{\it Ansatz} Local-Density Approximation (BALDA) 
in its fully numerical formulation~\cite{gao_PRB} (for odd $N$ the ground state has 
a finite polarization and thus a generalization to spin-BALDA is necessary).
The BALDA method allows us to efficiently treat complex systems with a very large number of sites.  
The combined use of DMRG and BALDA allows a detailed understanding of the 
problem and paves the way for further extensions.

{\it The pairing gap} --- 
We first discuss the tendency to pairing by analyzing the pair binding energy defined as 
\begin{equation}\label{eq:pair_binding}
E_{\rm P}=E_{\rm GS}(N+2)+E_{\rm GS}(N)-2\,E_{\rm GS}(N+1)
\end{equation}
where $E_{\rm GS}=\langle \Psi_{\rm GS}|{\hat {\cal H}}|\Psi_{\rm GS}\rangle$ is the ground-state energy. The pair binding energy of finite Hubbard chains with attractive interactions has been studied in Ref.~\onlinecite{marsiglio_1997}. In Table~\ref{table:one} we report results for $N=30$ fermions 
in a lattice with $L=100$ sites, inside a trap with $V_2/t=4\times 10^{-3}$. 
$E_{\rm P}$ is negative, signaling a tendency to pairing. 
\begin{table}
\caption{
	Ground-state and pair-binding energies for $N=30$, $L=100$, and $V_2/t=4\times 10^{-3}$. 
	The agreement between DMRG and BALDA for $E_{\rm GS}$ is quite satisfactory even for $U/t=-20$, 
	where the deviation is about $2.2 \%$. However, BALDA tends to overestimate $E_{\rm P}$ 
	even at moderate values of $U/t$. The ``$\times$" sign indicates that the spin-BALDA 
	code for $31$ atoms does not converge for $U/t=-20$.\label{table:one}}
\begin{center}
\begin{tabular}{ccccc}
	\hline
	\hline
	$U/t$ & $E^{\rm \scriptscriptstyle BALDA}_{\rm GS}/(t L)$ & 
	$E^{\rm \scriptscriptstyle DMRG}_{\rm GS}/(t L)$ & 
	$E^{\rm \scriptscriptstyle BALDA}_{\rm P}/t$ & $E^{\rm \scriptscriptstyle DMRG}_{\rm P}/t$\\\hline
	$-0.5$  & $-0.35824$ & $-0.35832$ &$-0.0283$ & $-0.0213$ \\\hline
	$-1$    & $-0.39336$ & $-0.39340$ &$-0.0614$ & $-0.0577$ \\\hline
	$-2$    & $-0.47672$ & $-0.47631$ &$-0.3265$ & $-0.2442$ \\\hline
	$-4$    & $-0.70693$ & $-0.69010$ &$-5.1008$ & $-1.3278$ \\\hline
	$-20$   & $-3.05320$ & $-2.98536$ &$\times$ & $-16.4217$\\
\hline\hline
\end{tabular}
\end{center}
\end{table}

\begin{figure}
\begin{center}
\tabcolsep=0cm
\begin{tabular}{c}
\includegraphics[width=0.70\linewidth]{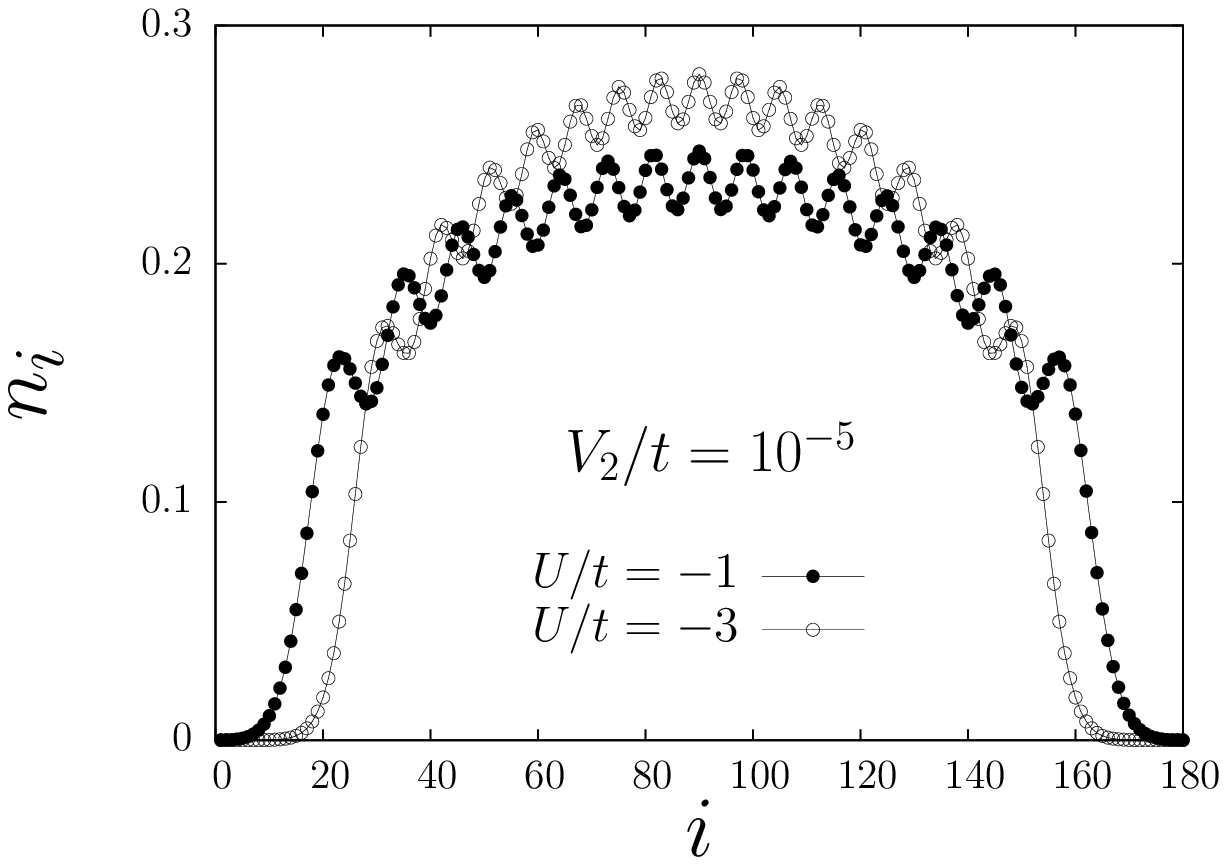}\\
\includegraphics[width=0.70\linewidth]{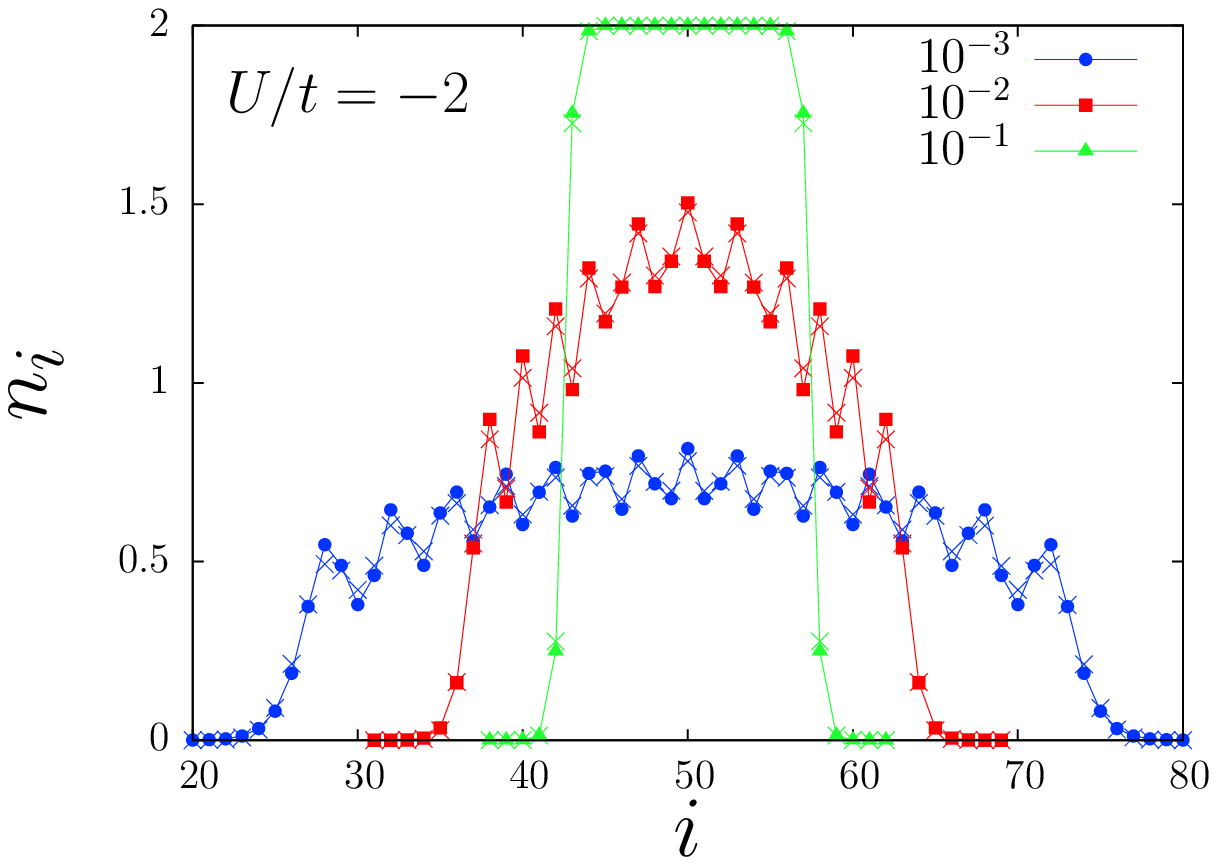}
\end{tabular}
\caption{(color online) Top panel: DMRG results for the site occupation $n_i$ as a function of site position $i$ 
	for a system with $N=30$ fermions in $L=180$ lattice sites, and in the presence of a harmonic potential with
	$V_2/t=10^{-5}$. For this value of $V_2/t$ BALDA overestimates the ADW amplitude. 
	Bottom panel: DMRG results (crosses) for $N=30$, $L=100$, and $U/t=-2$ are compared with BALDA data (filled symbols).
	$V_2/t$ is increased from $10^{-3}$ to $10^{-1}$. The thin solid lines are just a guide for the eye.\label{fig:one}}
\end{center}
\end{figure}

{\it Atomic-density waves} --- The pairing is associated with the presence of ADWs, 
which are stabilized by the harmonic potential. 
In Fig.~\ref{fig:one} we report our numerical results 
for the site occupation of a gas with $N=30$ atoms. The 
consequences of Luther-Emery pairing in the presence of confinement are dramatic. 
For $U<0$ the site occupation exhibits a density wave (with $N/2$ peaks in a weak trap), reflecting the tendency of atoms with different pseudospins to form stable spin-singlet dimers that are delocalized over the lattice. 

For small $V_2/t$ (see the top panel of Fig.~\ref{fig:one}) $n_i$ in the bulk of the trap ($80 \leq i \leq 100$) 
can be fitted to an ADW of the form 
$
n_i={\widetilde n}+A_{\rm \scriptscriptstyle ADW}\cos{(k_{\rm \scriptscriptstyle ADW}\,i+\varphi)}
$. 
For example, for $V_2/t=10^{-5}$ we find $k_{\rm \scriptscriptstyle ADW}=0.73$ for $U/t=-1$ and $k_{\rm \scriptscriptstyle ADW}=0.84$ for $U/t=-3$. In such a weak confinement the oscillations of the site occupation extend to regions far away from the center of the trap, where they are characterized by smaller edge wavenumbers. 
For $V_2=0$ bosonization predicts~\cite{giamarchi} an incipient ADW with wavenumber $2k_{\rm F}$, $k_{\rm F}$ being the Fermi wavenumber, quenched by strong quantum fluctuations. 
In the present case we find $k_{\rm \scriptscriptstyle ADW}\simeq 2k^{\rm eff}_{\rm F}$, where 
the effective Fermi wavenumber $k^{\rm eff}_{\rm F}=\pi {\widetilde n}/2$ is determined by the average 
density in the bulk of the trap (note that $k_{\rm \scriptscriptstyle ADW} = \pi$ when the average density in the bulk reaches half filling).

Finite-size effects become important on increasing $V_2$ (see the bottom panel of Fig.~\ref{fig:one}) 
and a simple fitting formula such as the one used above does not work even at the center 
of the trap. Eventually when $V_2/t\approx 10^{-1}$ a region of doubly-occupied sites develops at 
the center of the trap: spin-singlet dimers, which in a weak trap are delocalized, 
are squeezed close together to produce an extended region of $\approx N/2$ doubly-occupied consecutive sites.

\begin{figure}
\begin{center}
\tabcolsep=0cm
\begin{tabular}{cc}
\includegraphics[width=0.50\linewidth]{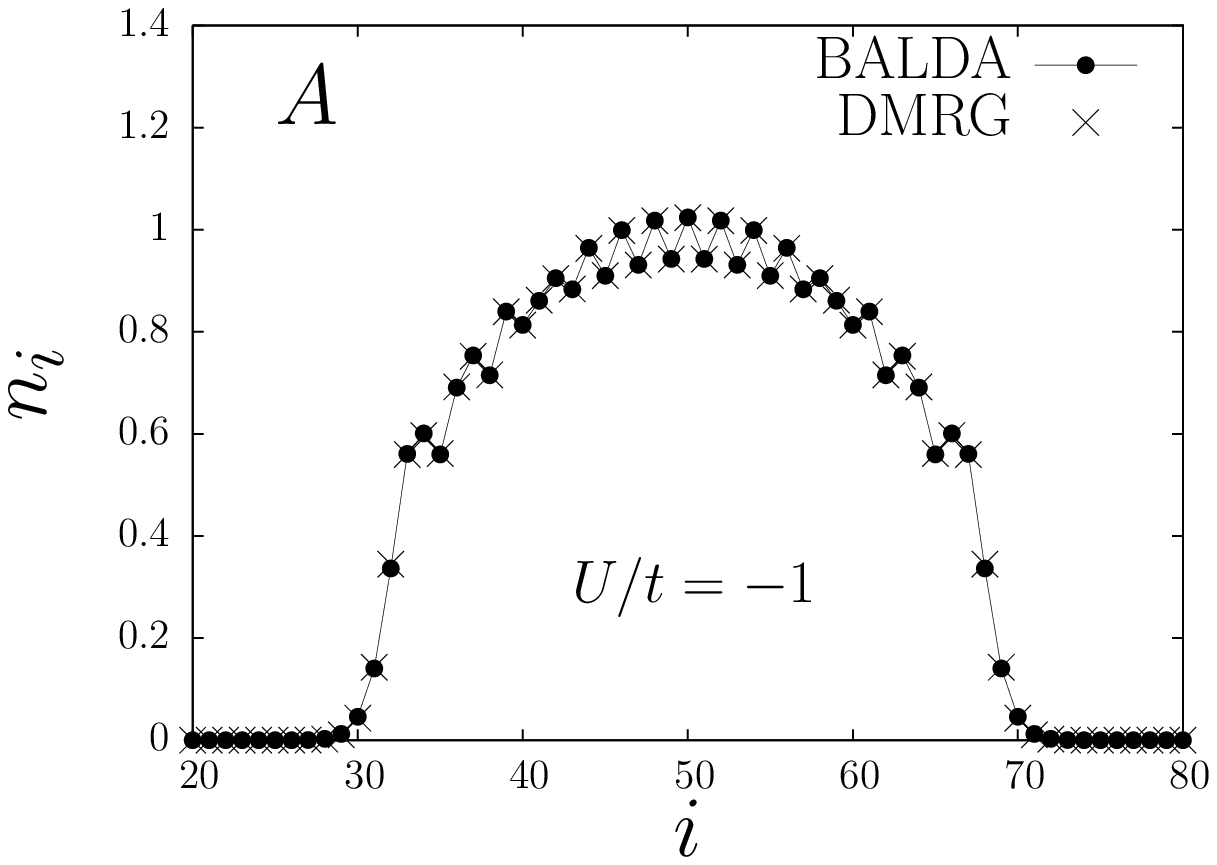}&
\includegraphics[width=0.50\linewidth]{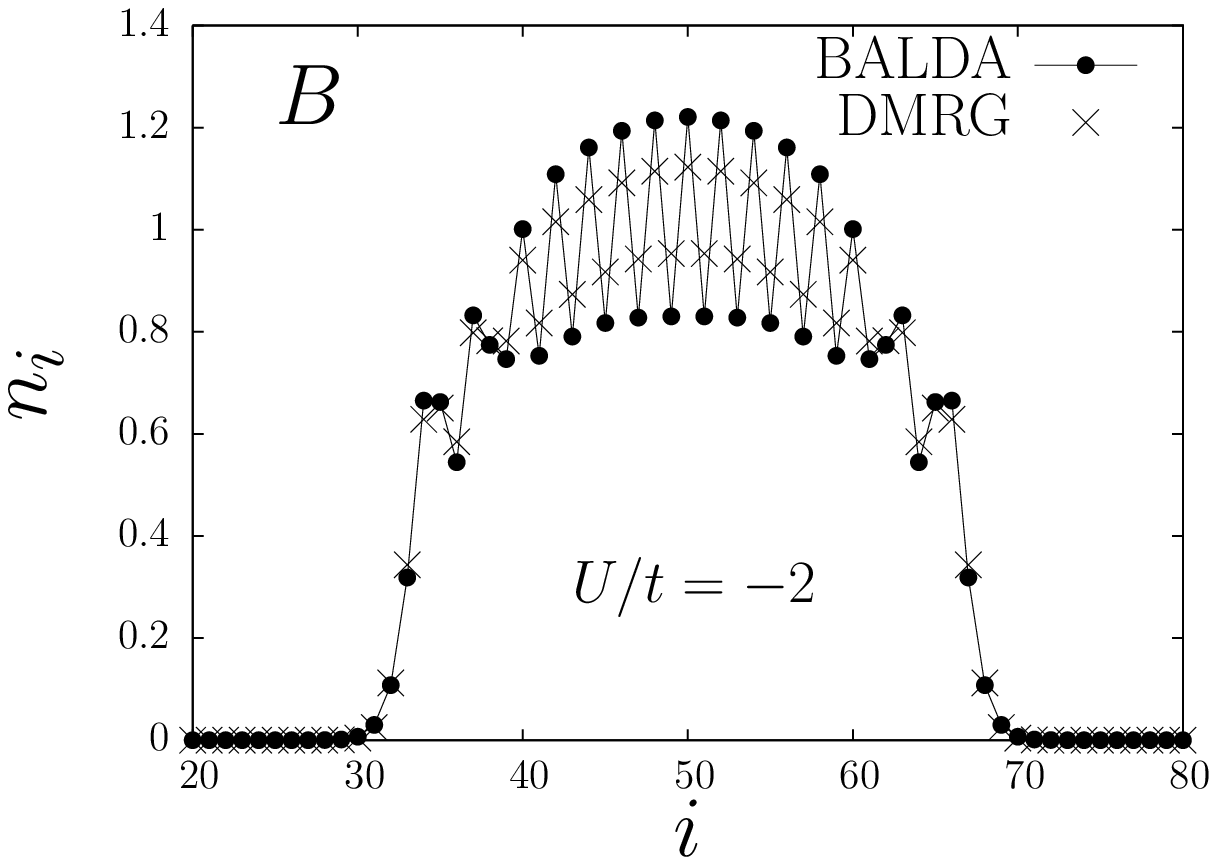}\\
\includegraphics[width=0.50\linewidth]{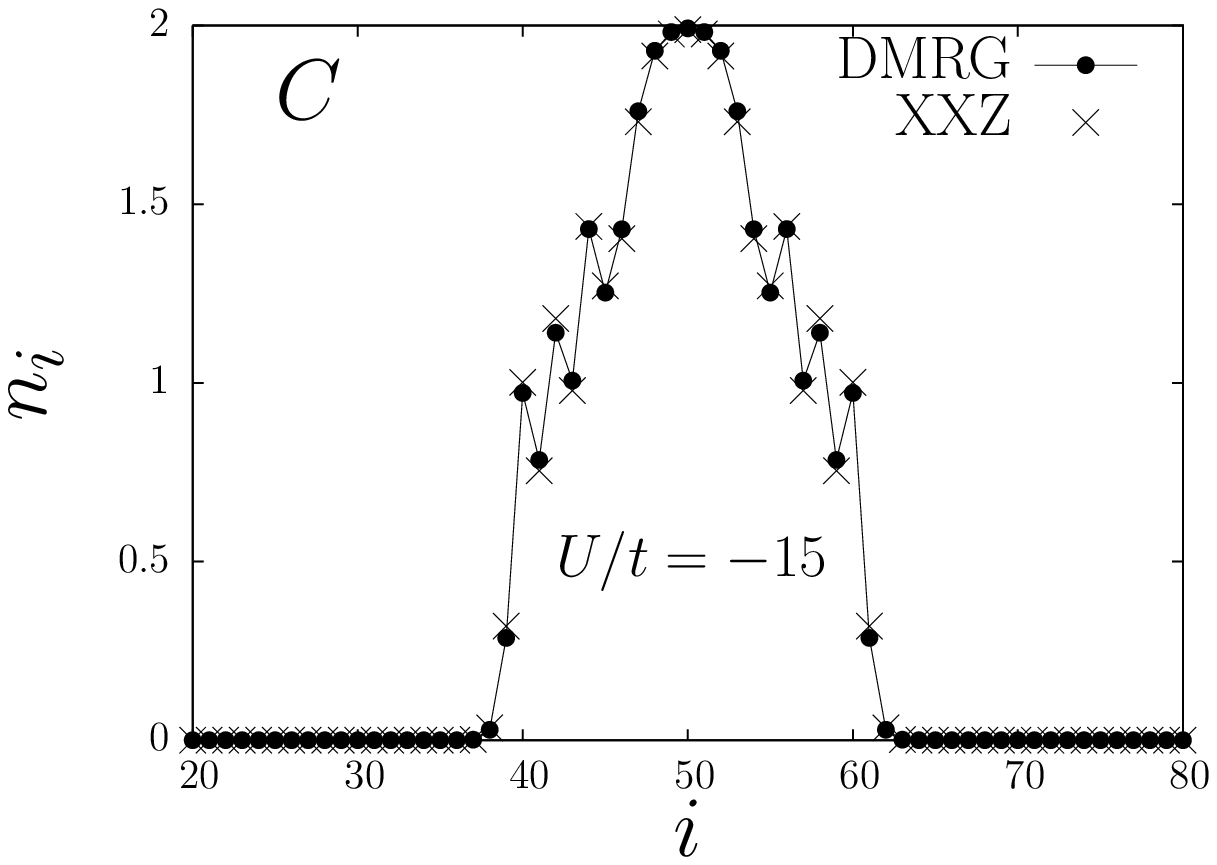}&
\includegraphics[width=0.50\linewidth]{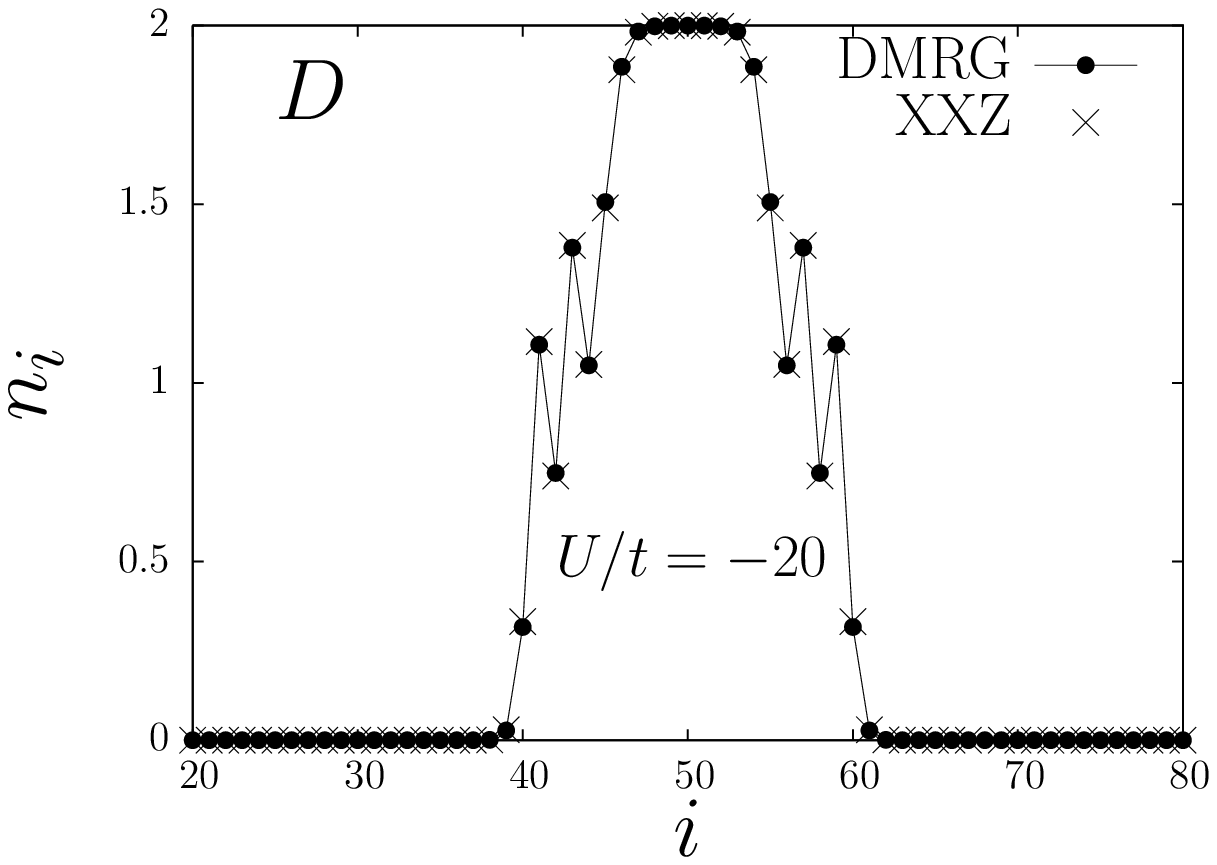}
\end{tabular}
\end{center}
\caption{Site occupation $n_i$ as a function of $i$ for $N=30$, 
	$L=100$, and  $V_2/t=4 \times 10^{-3}$. 
	Panels $A$ and $B$: DMRG results (crosses) are compared with BALDA data (filled circles). 
	Panels $C$ and $D$: ${\rm DMRG}$ results 
	for the Hamiltonian (\ref{eq:hubbard}) (filled circles) are 
	compared with ${\rm DMRG}$ results for the strong-coupling 
	Hamiltonian (\ref{eq:strong_coupling}) (crosses). The thin solid lines are just a guide for the eye.\label{fig:two}}
\end{figure}

In Fig.~\ref{fig:two} we show how the ADWs evolve with increasing $|U|/t$ 
at fixed $V_2/t$. For weak-to-intermediate coupling ADWs 
are present in the bulk of the trap. The agreement between the ${\rm BALDA}$ and the 
${\rm DMRG}$ results is  excellent for $|U|/t \leq 1$. With increasing $|U|/t$ the 
${\rm BALDA}$ scheme deteriorates~\cite{footnote}, leading to an overestimation of the 
amplitude of the ADWs (see panel $C$). According to ${\rm DMRG}$, the bulk 
ADWs disappear in the extreme strong-coupling limit (see panel $D$). For $|U|/t \gg 1$ 
a flat region of doubly-occupied sites emerges at the trap center, resembling 
that described above for the case of weak interactions and strong confinement (see the bottom panel of Fig.~\ref{fig:one}). 

The disappearance of the ADWs at strong coupling can be explained by mapping 
the Hamiltonian (\ref{eq:hubbard}) onto a spin-$1/2$ XXZ model~\cite{giamarchi,luther_emery},
\begin{eqnarray}
	\label{eq:strong_coupling}
	{\hat {\cal H}}_{\infty}&=&\sum_i\sum_{\alpha=x,y,z}
	J_\alpha {\hat \sigma}^\alpha_i{\hat \sigma}^\alpha_{i+1}+
	\sum_i B_i {\hat \sigma}^z_i\,,
\end{eqnarray}
with $J_x=J_y=-J_z=-t^2/|U|$ and $B_i=V_2(i-L/2)^2$. 
The total site occupation operator ${\hat n}_i$ is related to ${\hat \sigma}^z_i$ by
${\hat n}_i=1+{\hat \sigma}^z_i$. Particle-number conservation requires working in a sector 
with fixed total magnetization $\langle \sum_i{\hat \sigma}^{z}_i\rangle=N-L\equiv M$. 
In the limit $|U|/t\rightarrow \infty$, $J_\alpha$ is negligibly small and thus finding the 
ground state of ${\hat {\cal H}}_{\infty}$ is equivalent to solving the problem of orienting a 
collection of spins in a nonuniform magnetic field in order to minimize the Zeeman energy in producing a 
magnetization $M$. Thus, for $|U|/t\rightarrow \infty$ one expects a classical 
state with $\langle{\hat \sigma}^{z}_i\rangle=1$ ($\langle {\hat n}_i\rangle=2$) in $N/2$ 
sites at the trap center where $B_i$ is small, and $\langle{\hat \sigma}^{z}_i\rangle=-1$ 
($\langle {\hat n}_i\rangle=0$) in the remaining $L-N/2$ sites. 

{\it Observability of the ADWs}---
ADWs can be detected by a measurement of the elastic contribution to the light-scattering 
diffraction pattern, {\it i.e.} the Fraunhofer structure factor
\begin{equation}\label{eq:diffraction}
S_{\rm el}(k)=\frac{1}{N^2}|\sum_{j}e^{-ik j}n_j|^2\,,
\end{equation}
through the appearance of a peak at $k=k_{\rm \scriptscriptstyle ADW}$.

\begin{figure}
\begin{center}
\includegraphics[width=1.00\linewidth]{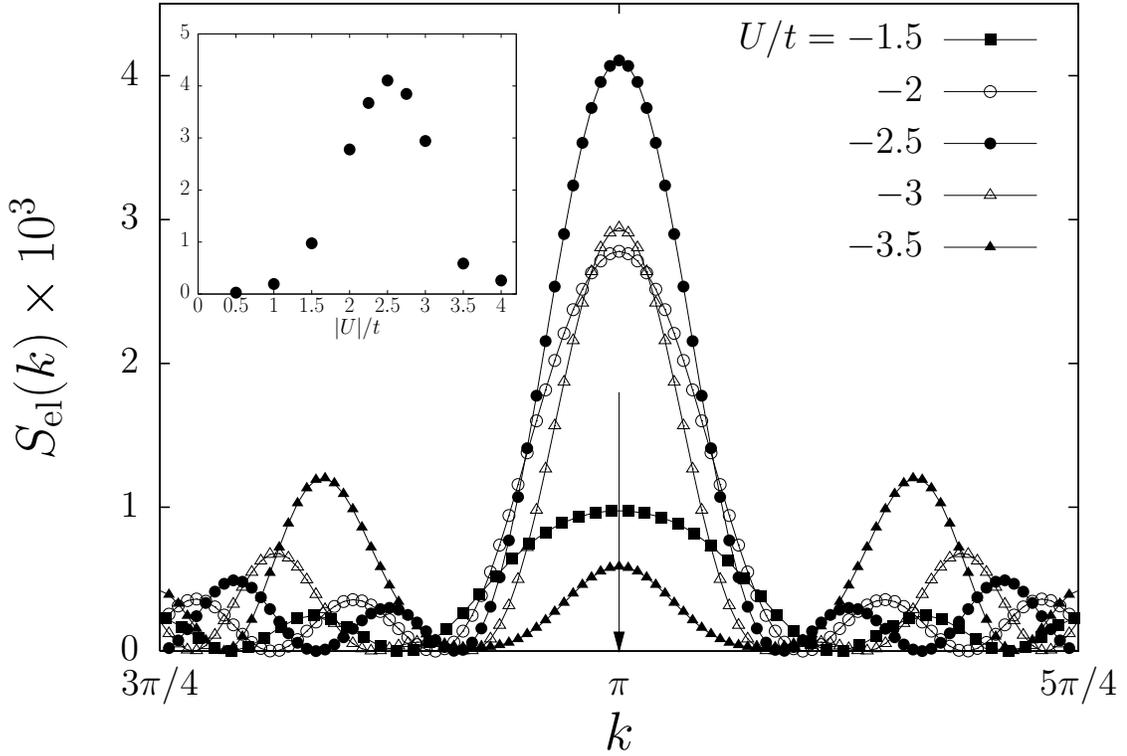}
\caption{
	DMRG data for $S_{\rm el}(k)$ as a function of $k$ 
	in the range $3\pi/4\leq k \leq 5\pi/4$, for $N=30$, $L=100$, 
	and $V_2/t=4 \times 10^{-3}$. The long vertical arrow indicates the value of 
	$k_{\rm \scriptscriptstyle ADW}$. In the inset we show $S_{\rm el}(\pi)$ 
	as a function of $|U|/t$.\label{fig:three}
	}
\end{center}
\end{figure}

In Fig.~\ref{fig:three} we show the most favorable situation where there is a 
wide region in the trap with oscillations at $k_{\rm \scriptscriptstyle ADW}=\pi$, 
inducing a well defined peak in $S_{\rm el}(k)$. The height of the peak is non-monotonic as a function of
$|U|/t$, as a consequence of the aforementioned crossover between the $U\rightarrow 0^-$ and $U\rightarrow -\infty$ limits, and is best observed when this ratio is of order two.
However, in a case such as that shown in the top panel of Fig.~\ref{fig:one}, where the density oscillations extend into regions far away from the trap center, the Fraunhofer structure factor peaks at a slightly lower wavenumber, $k_{\rm peak}\simeq 0.96k_{\rm \scriptscriptstyle ADW}$.

The main results of this work, {\it i.e.} negative pair binding energies and delocalized dimers forming ADWs in weak traps, are not a result of finite-size effects. We have carefully studied the scaling of these results in the thermodynamic limit following the procedure proposed in Ref.~\onlinecite{sachdev}, {\it i.e.} $N\rightarrow \infty$ and $V_2/t\rightarrow 0$ with $N\sqrt{V_2/t}={\rm const}$. We find that in this limit (i) $E_{\rm P}$ approaches a finite negative value and (ii) the amplitude of the ADWs (calculated after subtracting the smooth Thomas-Fermi site-occupation profile) approaches a finite value. For example, for $U/t=-2$ and $N^2 V_2/t=3.6$ we find $E_{\rm P}(N\gg 1)/t=-0.184-0.171\exp{(-N/28.61)}$ and $A_{\rm \scriptscriptstyle ADW}(N\gg 1)=0.032+0.114\exp{(-N/39.93)}$.

In summary, we have shown how the Luther-Emery phase emerges in 
an ultracold Fermi gas subject to parabolic trapping inside a $1D$ optical lattice.
We have demonstrated that the interplay between attractive interactions 
and harmonic confinement leads to coexistence of spin pairing and atomic-density waves. 
The existence of a finite pairing gap $\Delta=-E_{\rm P}/2$ can be tested 
{\it via} radio-frequency spectroscopy~\cite{rf}. 
Atomic-density waves can be detected from the elastic contribution to the light-scattering diffraction pattern. 
At strong coupling atomic-density waves change into a state 
in which spin-singlet dimers form an extended region of doubly occupied sites at the center of the trap.

\acknowledgments
This work was partially supported by FAPESP and CNPq. M.P. gratefully acknowledges useful discussions with M. Rigol. The DMRG calculations have been made using the DMRG code released
within the ``Powder with Power" Project (www.qti.sns.it). 

After this work was completed and submitted, we became aware of a related work of F. Karim Pour {\it et al.}~\cite{pour_august_2006}.


\begin{thebibliography}{99}
\bibitem{footnote} 
	This is different from the case $U>0$, 
	where the BALDA profiles agree with quantum Monte Carlo and DMRG data even 
	for $U/t>1$~\cite{gao_PRB,superlattice}. 
	Since the $U<0$ Bethe {\it Ansatz} is obtained from that at $U>0$ by means of an exact transformation, 
	this shows that the locality assumption inherent in the LDA is less well satisfied 
	for attractive than for repulsive interactions.	In fact, it appears that the ${\rm BALDA}$ performance 
	for a fixed value of $|U|/t$ depends on the value of $V_2/t$
\end{thebibliography}
\end{document}